\documentclass[portrait,11pt]{article}
\usepackage[sectionbib,round]{natbib}
\usepackage{times,amsmath,amsfonts}
\usepackage[dvips]{graphicx,graphics}

\usepackage{amsfonts,amsmath,amssymb}
\usepackage{url}
\usepackage{hyperref}
\hypersetup{colorlinks,%
citecolor=black,%
filecolor=black,%
linkcolor=blue,%
urlcolor=blue,%
pdftex}

\newcommand{\bqn}{\begin{eqnarray*}}
\newcommand{\eqn}{\end{eqnarray*}}
\newcommand{\bq}{\begin{eqnarray}}
\newcommand{\eq}{\end{eqnarray}}

\begin{document}
\pagenumbering{gobble}

\title{Introduction to Brain and Medical Images}
\author{Moo K. Chung \\
University of Wisconsin-Madison\\
\quad \tt{mkchung@wisc.edu}}
\maketitle

This article is based on the first chapter of book \citet{chung.2013.SCM}, where brain and medical images are introduced. The most widely used brain imaging modalities are magnetic resonance images (MRI), functional-MRI (fMRI) and diffusion tensor images (DTI). A brief introduction to each imaging modality is explained. Further, we explain what kind of curve, volume and surface data that can be extracted from each modality. 

\section{Introduction}

MRI depends on the response of magnetic fields in generating digital images that provide structural information about brain noninvasively. Compared to the computed tomography (CT), MRI has been mainly used for {\em in vivo} imaging of the brain due to higher image contrast in soft tissues.  Unlike CT, MRI does not use X-ray for imaging so there is no risk of radiation exposure. MRI produces images based on the spin-lattice relaxation time (T1), the spin-spin relaxation time (T2) and the proton density ($\rho$) \citep{bernstein.2005}. The widely used T1- and T2-weighted imaging weight the contribution of one component and minimize the effect of the other two. The T1-weighted MRI is more often used in anatomical studies compared to the T2-weighted MRI. Structural images you are seeing are most likely T1-weighted MRI.

DTI also uses MRI scanners but it specifically enables the measurement of the diffusion of water molecules in the brain tissues. At each voxel in DTI,  
a rate of diffusion and a preferred direction of diffusion are encoded as $3 \times 3$ symmetric and positive definite matrix called diffusion tensor. In each voxel, there are countless number of neural fibers. So the direction of diffusion follows neural fiber tracts in an average sense within each voxel. By tracing the diffusion direction via tractography, it is possible to represent the white matter fiber tracts visually. fMRI also uses MRI scanners in measuring brain activity by detecting changes in blood flow using the blood-oxygen-level-dependent (BOLD) contrast. For last two decades, fMRI has been dominating brain mapping research since it is easy to scan and there is no exposure to radiation. Compared to MRI, fMRI is more frequently contaminated by various types of noise.  In order to obtain the underlying signal, various advanced statistical procedures are used. The resulting statistical maps are usually color coded to show the strength of activation in the whole brain. 

In this article, various brain image data types obtained from different imaging modalities will be introduced with simple {\tt MATLAB} functions for image processing and visualization. 
The most difficult part for statisticians and new researchers trying to analyze brain image data is getting the data into a computer programming environment. We also briefly go over manipulating volume and surface data.   The {\tt MATLAB}  code and sample data used in this chapter can be downloaded from 
\url{http://brainimaging.waisman.wisc.edu/~chung/BIA}.

\section{Image Volume Data}

\begin{figure}
\centering
\includegraphics[width=1\linewidth]{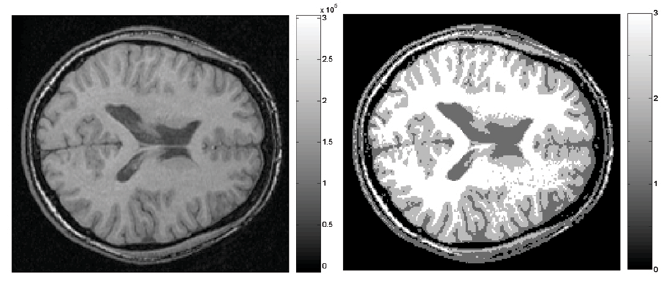}
 \caption{Left; the axial cross-section of MRI. Right: Tissue segmentation into three different classes (white and gray matter and CSF) using the neural network classifier \citep{kollakian.1996}.}\label{fig:chap1.segmentation}
\end{figure}

The brain MRI is a 3D array $\mathcal{I}({\bf x})$ containing tissue intensity values at voxel position ${\bf x} \in \mathbb{R}^3$ 
Figure \ref{fig:chap1.segmentation} shows  a usual example of  axial cross-section of MRI. We usually do not analyze tissue intensity values directly in quantifying MRI. MRIs go through various image processing steps such as intensity normalization and tissue segmentation before any quantification is attempted. Image intensity normalization is required to minimize the intensity variations within the same tissue that are caused by magnetic field distortion. 

Each voxel can be classified into three different tissue types: cerebrospinal fluid (CSF), grey matter and white matter mainly based on image intensity values. A neural network classifier \citep{kollakian.1996} or a Gaussian mixture model \citep{ashburner.2000} have been used for the
classification. Figure \ref{fig:chap1.segmentation} shows the result obtained by a neural network classifier \citep{kollakian.1996}. The collection of identically classified voxels has been mainly used in voxel-based morphometry (VBM), which will be discussed in detail in a later chapter. VBM as implemented in the statistical parametric mapping (SPM) package 
(\url{www.fil.ion.ucl.ac.uk/spm}), is a fully automated image analysis technique allowing identification of regional differences in gray matter (GM) and white matter (WM) between populations without a prior region of interest. VBM starts with normalizing each structural MRI to the standard SPM template and segmenting it into white and gray matter and cerebrospinal fluid (CSF) based on a Gaussian mixture model \citep{ashburner.1997,ashburner.2000}. Based on a prior probability of each voxel being the specific tissue type, an iterative Bayesian approach is used to get an improved estimate of the posterior probability. This probability is usually referred to as the gray or white matter density. Afterwards the density maps are warped into a normalized space and compared across subjects. VBM has been applied in cross-sectional studies on various anatomical studies: normal development \citep{good.2001,paus.1999}, autism \citep{chung.2004.ni}, depression \citep{pizzagalli.2004}, epilepsy \citep{mcmillan.2004} and mild cognitive impairment (MCI) \citep{johnson.2004}. 

\begin{figure}
\centering
\includegraphics[width=1\linewidth]{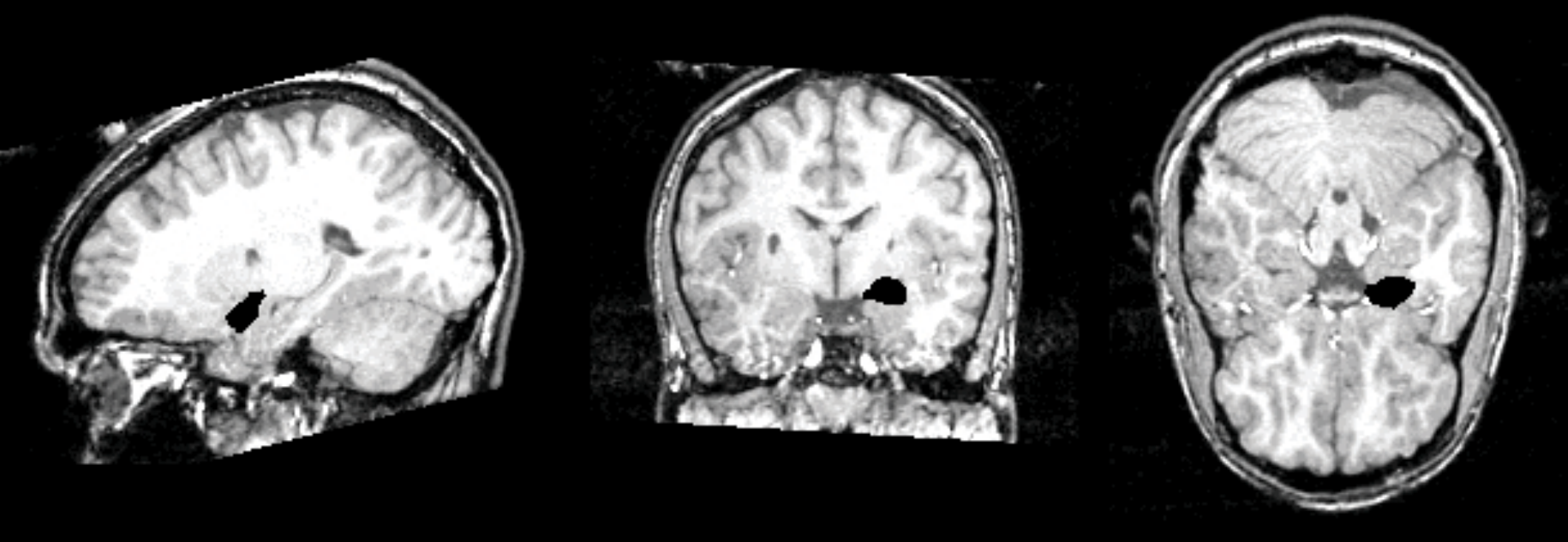}
 \caption{The manual segmentation of left amygdala at the mid-sagittal, coronal and axial cross-sections.  The amygdala was segmented using a prior information of adjacent structures such as anterior commissure and hippocampus \citep{nacewicz.2006}.}
\label{fig:chap1.amyg}
\end{figure}

The most widely used 3D MRI volume image format has been the Analyze 7.5 file format developed by the Biomedical Imaging Resource at Mayo Clinic for Analyze package (\url{www.mayo.edu/bir/Software/Analyze/Analyze.html}). An Analyze format consists of two files: an image  binary file ({\tt *.img}) and a header file ({\tt *.hdr}) that contains information about the binary image file. Various brain image analysis software and {\tt MATLAB} can read the Analyze 7.5 file format. 
\index{file formats ! Analyze}

Figure \ref{fig:chap1.amyg} shows an example of MRI and the manual segmentation of the left amygdala, which is colored in black. This manual segmentation is stored as a binary image consisting of 0's and 1's. In {\tt MATLAB}, to read the header file, we invoke the command  

\begin{verbatim}
>> header=analyze75info('left_amygdala.hdr');

header = 

         Filename: 'left_amygdala.hdr'
         FileModDate: '21-Sep-2003 15:00:02'
         ColorType: 'grayscale'
         Dimensions: [191 236 171 1]
         VoxelUnits: '  mm'
\end{verbatim}

The structure array {\tt header} contains various image information and can be accessed by calling, for instance, {\tt header.Dimensions} which shows the dimension of array {\tt  [191 236 171 1]}. The last dimension is reserved for temporally changing 3D volume images such as fMRI but it is not used for this example. To extract the amygdala segmentation and saved into a 3D array {\tt vol} of size $191 \times 236 \times 171$, we run

\begin{verbatim}
>>vol=analyze75read(header);
\end{verbatim}

{\tt vol} is then a 3D binary array of a left amygdala consisting of zeros and ones (Figure \ref{fig:chap1.amyg}). The array {\tt vol} consists of all zeros except the voxels where amygdala is defined. 

\subsection{Amygdala Volume Data}

This is the subset of the 44 subject data set first published in \citet{chung.2010.ni} consisting of the both left and right amygdala binary volume of 22 subjects. There are 22 normal control subjects and 24 autistic subjects. The data set is stored as {\tt amygdala.volume.data} containing variables age, total brain volume ({\tt brain}), eye fixation duration ({\tt eye}), face fixation duration ({\tt face}), group variable indicating autistic or normal control ({\tt group}), left amygdala volume ({\tt leftvol}) and right amygdala volume ({\tt rightvol}) data. The main scientific hypothesis of interest is to quantify volume and  shape differences between the groups. In this section, we will show how to load the data set and do a simple volume computation. Since the image resolution is $1 \times 1 \times 1$ mm, the amygdala volume is simply computed by counting the number of voxels belonging to the amygdala. 

\begin{verbatim}
load amygdala.volume.mat

left=zeros(22,1);
right=zeros(22,1);
for i=1:22
    left(i)=sum(sum(sum(squeeze(leftvol(i,:,:,:)))));
    right(i)=sum(sum(sum(squeeze(rightvol(i,:,:,:)))));
end;
\end{verbatim}
\index{volume!amygdala}\index{amygdala}

\begin{figure}[t]
\begin{center}
\includegraphics[width=1\linewidth]{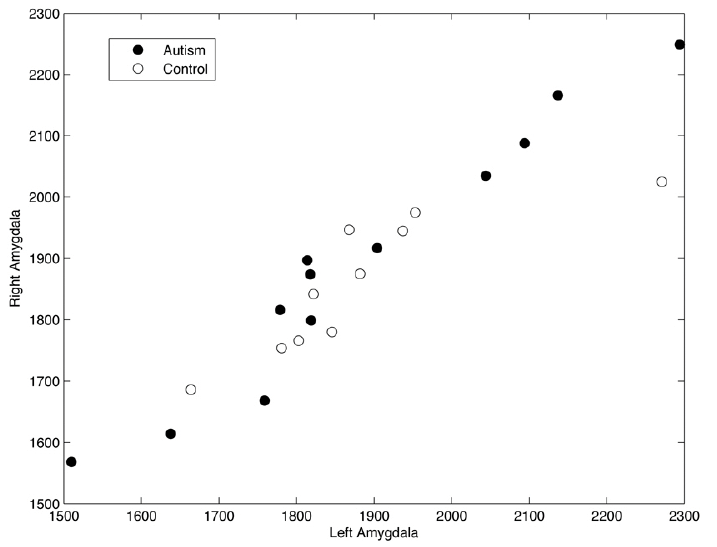}
 \caption{The plot showing the left and right amygdala volumes of autistic (black) and control (white) subjects. There is no visible group differences in the plot. The volume based method is not effective in discriminating between the groups.}\label{fig:chap1.volumeplot}
 \end{center}
\end{figure}

The {\tt group} variable is 1 for autistic and 0 for normal controls. The left amygdala volume is computed and stored as {\tt al} for instance. The amygdala volumes for all subjects are displayed in Figure \ref{fig:chap1.volumeplot} showing no clear separation of the groups. 
 
\begin{verbatim}
al= left(find(group));             
ar=right(find(group));
cl= left(find(~group));
cr=right(find(~group));

figure;
plot(al,ar,'or', 'MarkerEdgeColor','k', ...
'MarkerFaceColor','k', 'MarkerSize',7)
hold on
plot(cl,cr,'ob', 'MarkerEdgeColor','k', ...
'MarkerFaceColor','w', 'MarkerSize',7)
legend('Autism','Control')
xlabel('Left Amygdala')
ylabel('Right Amygdala')
set(gcf,'Color','w');
\end{verbatim}

A more advanced technique such as weighted spherical harmonic representation \citep{chung.2007.TMI,chung.2010.ni}  is needed to increase the discrimination power.

\section{Surface Mesh Data}
\index{surface mesh}
Simple 3D neuroanatomical objects like hippocampus and amygdala can be represented by their 2D boundary as triangle meshes using  the  marching cubes algorithm \citep{lorensen.1987}, which is implemented in {\tt MATLAB}.

To extract the surface mesh out of {\tt vol}, we run

\begin{verbatim}
>>surf = isosurface(vol)

surf = 

    vertices: [1270x3 double]
       faces: [2536x3 double]
\end{verbatim}

\begin{figure}
\begin{center}
\includegraphics[width=1\linewidth]{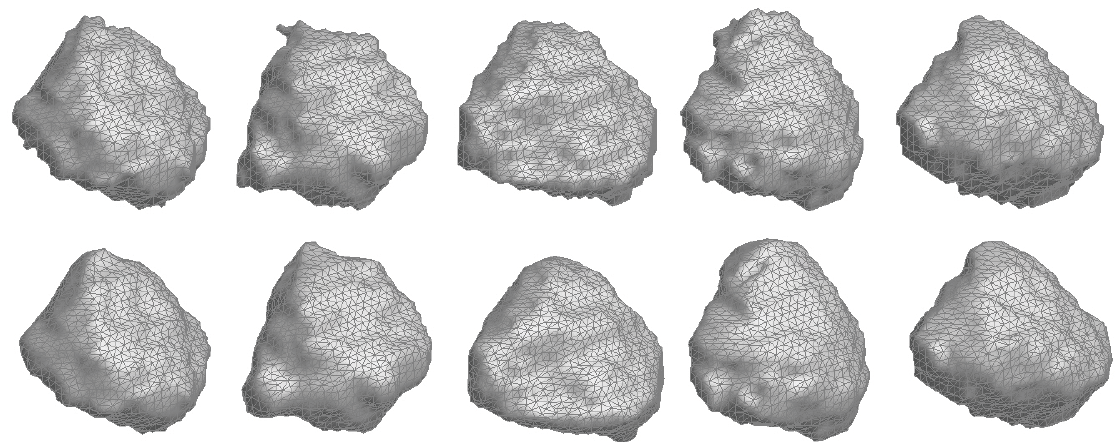}
 \caption{Top: triangle mesh representation of five different left amygdala surfaces. The mesh noises such as sharp peaks need to be smoothed out using surface smoothing technique. Here we used the weighted-SPHARM representation \citep{chung.2010.ni}, which will be studied in a later chapter.}\label{fig:chap1.amygmesh}
 \end{center}
\end{figure}

The {\tt MATLAB} function {\tt isorsurface} extracts the boundary of the amygdala and represents it as the structure array {\tt surf} consisting of 1270 vertices and 2563 triangles. The visualization can be done with
\begin{verbatim}
>>figure_wire(surf,'yellow')
\end{verbatim}
resulting in a surface mesh like Figure \ref{fig:chap1.amygmesh}.

More complex neuroanatomical objects like cortical surfaces are difficult to represent with such a simple algorithm. Substantial research have been done on extracting cortical surfaces from MRI \citep{chung.2003.ni,davatzikos.1995,macdonald.2000}. The human cerebral cortex has the topology of a 2D highly convoluted grey matter shell with an average thickness of 3mm. The outer boundary of the shell is called  the {\em outer cortical surface} while the inner boundary is called the {\em inner cortical surface}. The outer cortical surface is the boundary between the cerebrospinal fluid (CSF) and the gray matter while the inner cortical surface is the boundary between the gray and the white matters. Cortical surfaces are segmented from MRI using mainly a deformable surface algorithm and represented as a triangle mesh consisting of more than 40,000 vertices and 80,000 triangle elements  \citep{chung.2003.ni,macdonald.2000}. We assume cortical surfaces to be smooth 2D Riemannian manifolds topologically equivalent to a unit sphere \citep{davatzikos.1995}. The triangle mesh format contains information about vertex indices, the Cartesian coordinates of the vertices and the connectivity that tells which three vertices form a triangle.

\subsection{Topology of Surface Data} 
Surface extraction from MRI may cause possible topological defects such as  handles and bridges forming on brain surfaces. So it is crucial to able to  determine topological defects. Determining topological defects on the surface can be done by checking the topological invariants such as the Betti numbers or Euler characteristics.

Let $\chi$ be the Euler characteristic of a surface. The sphere has a Euler characteristic of 2. If the surface has genus $g$, the number of handles, the Euler characteristic is given by 
$$
\chi = 2-2g.
$$
Each handle in the object reduces $\chi$ by 2 while the increase of the disconnected components raise $\chi$ by 2. On the surface mesh, the Euler characteristic is given in terms of the number of vertices $V$,  the number of edges $E$ and  the number of faces $F$ using the polyhedral formula:
$$
\chi = V - E + F.
$$
\index{Euler characteristics}

\begin{figure}[t]
\begin{center}
\includegraphics[width=1\linewidth]{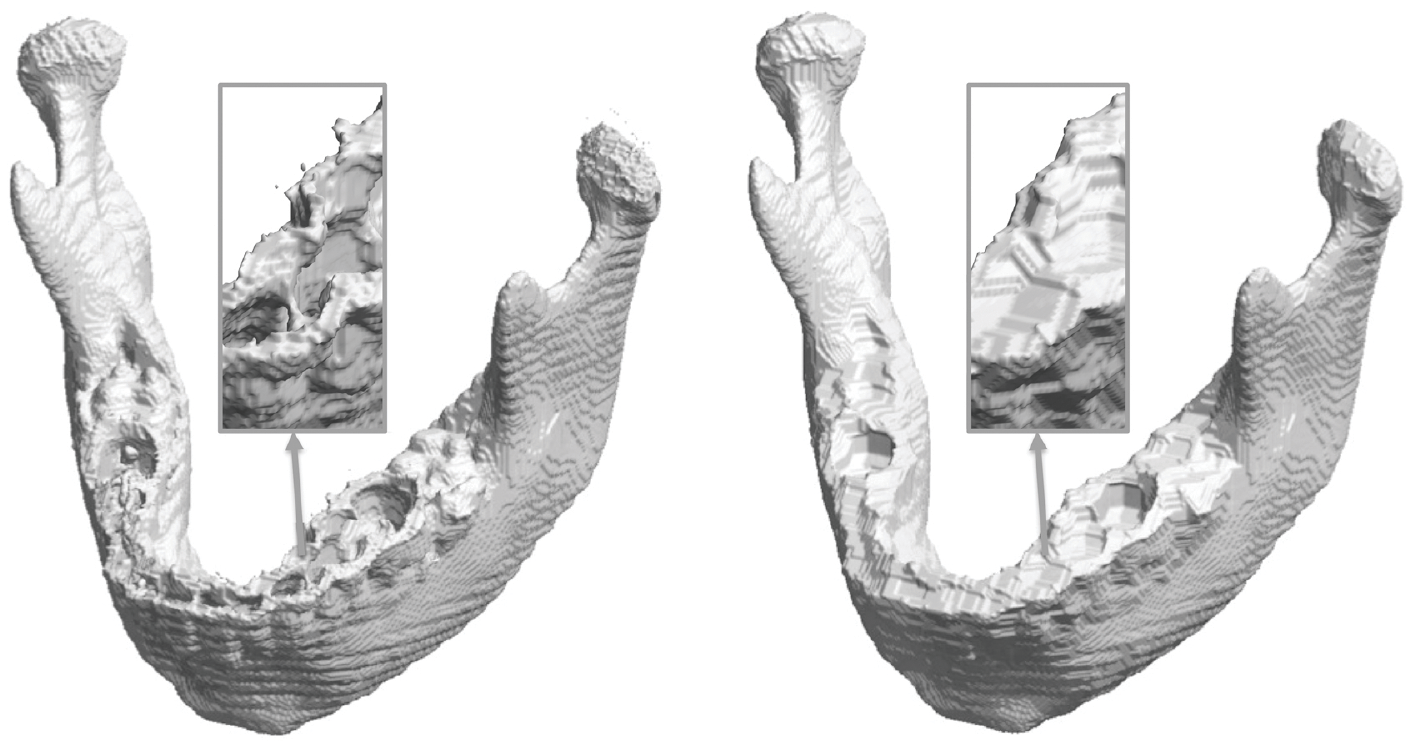}
\caption{Topological defects occur in any type of medical image segmentation. Bridges and handles are visible in the teeth regions of the mandible segmentation obtained from CT images. Topological correction is made on mandible binary segmentation and surface. Disjointed tiny speckles of noisy components are removed by labeling the largest connected component, and holes and handles are removed by the morphological closing operation.}
\label{fig:holepatching}
\end{center}
\end{figure}

Note that for each triangle, there are three edges. For a closed surface topologically equivalent to a sphere, two adjacent triangles share the same edge. Hence, the total number of edges is $E = 3F/2$. The relationship between the number of vertices and the triangles is $F=2V-4$. We simply need to compute the Euler characteristic as $\chi = V - F/2$ and check if it is 2 at the end.  All binary volumes produced the topologically correct surfaces without an exception. Fig.~\ref{fig:holepatching} shows an example of before and after the topology correction.

For any type of cortical surface mesh, if $V$ is the number of vertices, $E$ is the number of edges, and $F$ is the number of faces or triangles in the mesh, the Euler characteristic $\chi$ of the mesh should be constant, i.e. $\chi =V-E +F = 2$. Note that for each triangle, there are three edges. Since two adjacent triangles share the same edge, the total number of edges is $E = 3F/2$. Hence,  the relationship between the number of vertices and the triangles is $F=2V-4$. In the sample surface, we have 40,962 vertices and 81,920 triangles.

The Euler characteristic computation in brain imaging has been traditionally done in connection with correcting for topological defects in segmented anatomical objects \citep{segonne.2007, shattuck.2001}. For instance, the human cerebral cortex has the topology of a 2D highly convoluted grey matter shell with an average thickness of 3mm. The outer and inner boundaries are assumed to be topologically equivalent to a sphere  \citep{davatzikos.1995, macdonald.2000}. Image acquisition and processing artifacts, and partial voluming will likely produce topological defects such as holes and handles in cortical segmentation. 
Since it is also needed to remove the brain stem and other parts of the brain in the triangulated mesh representation of the cortex, the resulting cortical surfaces are not likely to have spherical topology. So it is necessary to automatically determine and correct the topological defects using the topological invariants such as the Euler characteristic.

Topological defects are often encountered in image reconstruction. For instance, since the mandible and teeth have relatively low density in CT, unwanted cavities, holes and handles can be introduced in mandible segmentation \citep{andresen.2000.TMI}. An example is shown in Figure \ref{fig:holepatching} where the tooth cavity forms a bridge over the mandible. In mandibles, these topological noises can appear in thin or cancellous bone, such as in the condylar head and posterior palate \citep{stratemann.2010}. If we apply the isosurface extraction on the topologically defect segmentation results, the resulting surface will have many tiny handles \citep{wood.2004,yotter.2009.MICCAI}. These handles complicate subsequent mesh operations such as smoothing and parameterization. So it is necessary to correct the topology by filling the holes and removing handles. If we correct such topological defects, it is expected the resulting isosurface is topologically equivalent to a sphere. 
There have been various topological correction techniques proposed in medial image processing. Rather than attempting to repair the topological defects of the already extracted surfaces \citep{wood.2004,yotter.2009.MICCAI}, we can perform  the topological simplification on the volume representation directly using morphological operations \citep{guskov.2001,yotter.2009.MICCAI}. The direct correction on surface meshes can possibly cause surfaces to intersect with each other \citep{wood.2004}. 

\subsection{Amygdala Surface Data}
\index{surface mesh! amygdala}

\begin{figure}[t]
\begin{center}
\includegraphics[width=1\linewidth]{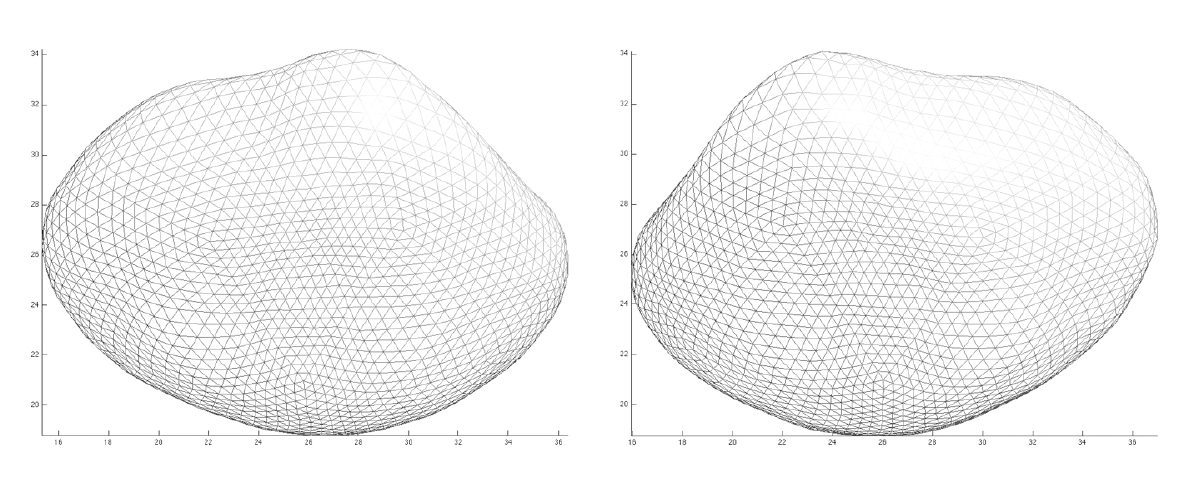}
\caption{Left and right amygdala surface template constructed by averaging surface coordinates in the data set {\tt chung.2010.NI.mat}.}
\label{fig:amygdalatemplates}
\end{center}
\end{figure}

This is the data set first published in \citep{chung.2010.ni} consisting of the both left and right amygdala surfaces of 44 subjects. There are 22 normal control subjects and 24 autistic subjects. All the amygdala surfaces went through image processing and meshes consist of  2562 vertices and 5120 faces. Mesh vertices anatomically match across subjects. So we can simply average the corresponding vertex coordinates to obtain the group average (Figure \ref{fig:amygdalatemplates}). 

\begin{verbatim}
load chung.2010.NI.mat

left_template.vertices = squeeze(mean(left_surf,1));
left_template.faces=sphere.faces;
figure; figure_wire(left_template,'yellow','white');

right_template.vertices = squeeze(mean(right_surf,1));
right_template.faces=sphere.faces;
figure; figure_wire(right_template,'yellow','white');
\end{verbatim}

\begin{figure}[t]
\begin{center}
\includegraphics[width=1\linewidth]{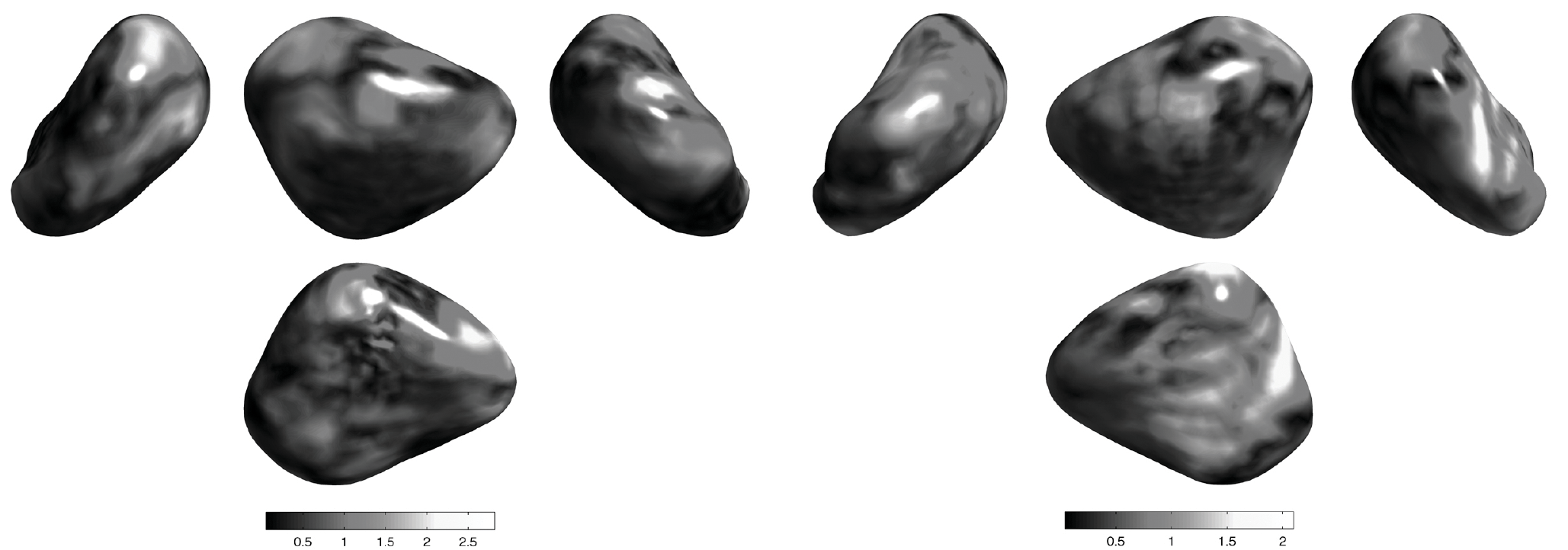}
\caption{Displacement that is required to move the template surface to the first subject is displayed in the origami  representation \citep{chung.2010.ni}. The unit is in mm showing a fairly small deformation scale.}
\label{fig:amygdalatemplate}
\end{center}
\end{figure}

The displacement vector field of warping the template surface to the 46 subject surfaces is given in $46 \times 2562 \times 3$ matrix. The displacement length of the first subject is shown in the origami representation introduced in \citep{chung.2010.ni} (Figure \ref{fig:amygdalatemplate}).
\index{displacement}
\index{vector field!displacement}

\begin{verbatim}

temp=reshape(left_template.vertices, 1, 2562,3);
temp=repmat(temp, [46 1 1]);
left_disp= left_surf - temp;

temp=reshape(right_template.vertices, 1, 2562,3);
temp=repmat(temp, [46 1 1]);
right_disp= right_surf - temp;

left_length= sqrt(sum(left_disp.^2,3));
figure_origami(left_template, left_length(2,:))

right_length= sqrt(sum(right_disp.^2,3));
figure_origami(right_template, right_length(2,:))
\end{verbatim}

\section{Landmark Data}
\index{landmark}
Either manual or automatic landmarks that identify the important regions and features of anatomy are often used in landmark-based morphometrics. Landmarks are also used in aligning anatomy using affine registration. The landmark data introduced in this section is manually identified along the mandible surfaces extracted from CT images. The CT images were obtained from GE multi-slice helical CT scanners. CT scans were converted to DICOM format and subsequently Analyze $8.1$ software package (AnalyzeDirect, Inc., Overland Park, KS) was used in segmenting binary mandible structure based on histogram thresholding. Then 24 landmarks are manually identified by an expert for two mandible surfaces. The landmarks across different surfaces anatomically correspond.  

\begin{figure}
\centering
\includegraphics[width=0.9\linewidth]{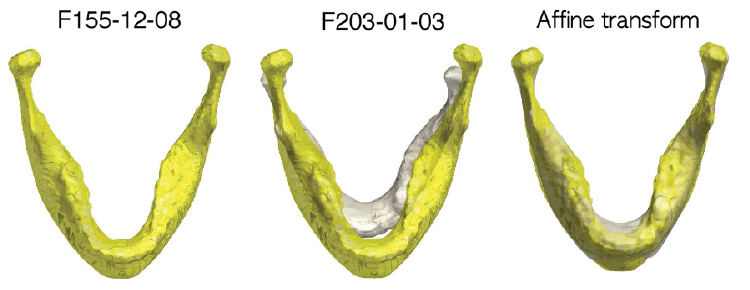}
\caption{Mandible F155-12-08 (yellow) is used as a fixed template and other mandibles are affinely aligned to F155-12-08. For example, smaller F203-01-03 (gray) is aligned to larger F203-01-03 by affinely enlarging it.} 
\label{fig:affine-alignment}
\includegraphics[width=0.8\linewidth]{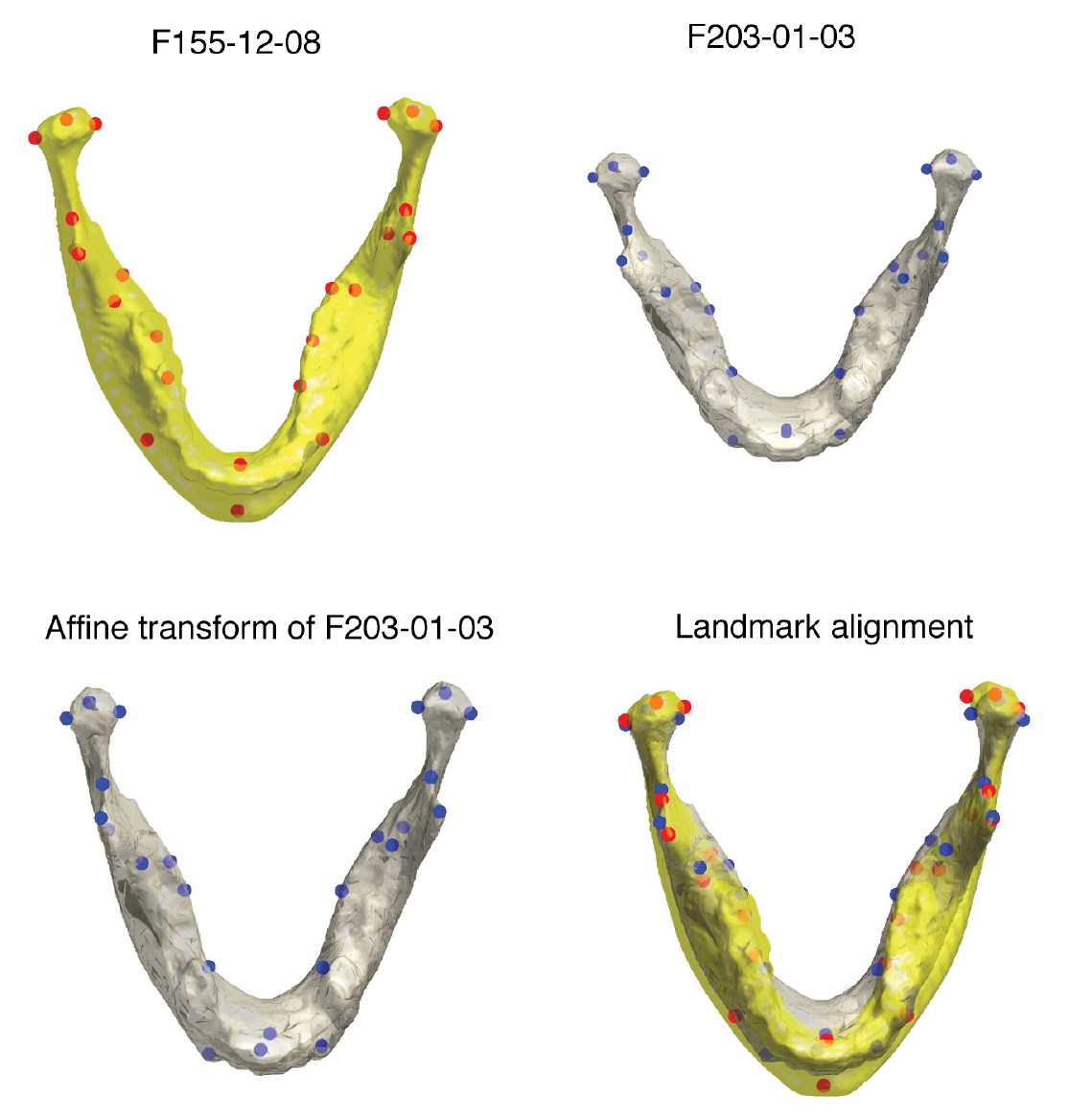}
\caption{Mandible F155-12-08 (yellow) is used as a template and mandibles F203-01-03 is affinely aligned to F155-12-08 by matching 24 manually identified landmarks. The affine transform does not exactly match the landmarks perfectly but minimizes the distance between them in a least squares fashion.} 
\label{fig:affine-alignment2}
\end{figure}

The mandible surfaces and the landmark data are stored in \\
{\tt mandible-landmarks.mat}. In the {\tt mat} file, {\tt id} contains the subject identifiers. {\tt id(44)} is the subject 
F155-12-08 while {\tt id(1)} is the subject F203-01-03. The mandible surfaces for these two subjects are shown in  Figure \ref{fig:affine-alignment}. Our aim is to align them affinely as close as possible. 
 
\begin{verbatim}
load mandible-landmarks.mat

figure; subplot(1,2,1);
hold on; figure_patch(F155_12_08, 'y', 0.5);
q=landmarks(:,:,44);
C=20*ones(24,1);
hold on; scatter3(q(:,1),q(:,2),q(:,3), C, C, 'filled', 'r')
view([90 60]);  zoom(1.2); camlight; 
title(id(44))

subplot(1,2,2);
figure_patch(F203_01_03, [0.74 0.71 0.61], 0.5);
p=landmarks(:,:,1);
C=20*ones(24,1);
hold on; scatter3(p(:,1),p(:,2),p(:,3), C, C, 'filled', 'b')
view([90 60]);  zoom(1.2); camlight; 
title(id(1))
\end{verbatim}

\subsection{Affine Transforms}
\index{affine transforms}

Anatomical objects extracted from 3D medical images are aligned using affine transformations to remove the global size differences. Figure \ref{fig:affine-alignment} illustrates an example where one mandible surface is aligned to another via an affine transform. 

The affine transform $T$ of point $p=(p_1, \cdots, p_d)' \in \mathbb{R}^d$ to $q=(q_1, \cdots, q_d)'$ is given by
$$q = Rp + c,$$
where the matrix $R$ corresponds to rotation, scaling and shear while $c$ corresponds to translation. Note that the affine transform is nonlinear. Note that
\bq T(ap+ bq) &=& R(ap+bq) + c\\
&=& a(Rp + c) + b(Rq + c) -c(a+b -1)\\
&=& aT(p) + bT(q) - c(a+b-1).\eq
Unless $a+b \neq 1$, the affine transform is not linear. However, the affine transform can be easily made into a linear form by augmenting the transform. The affine transform can be rewritten in a matrix form as
\bqn
\left(
\begin{array}{c}
 q   \\
1   \\
\end{array}
\right)
=
\left(
\begin{array}{cc}
 R &  c    \\
 0 \cdots 0&   1   \\
\end{array}
\right) \left(
\begin{array}{c}
 p   \\
1   \\
\end{array}
\right)
\label{affine-matrixform}.
\eqn
The affine transformation matrix $A$ is then given by 
$$A= \left(
\begin{array}{cc}
 R &  c    \\
 0 \cdots 0&   1   \\
\end{array}
\right).$$
Trivially $A$ is linear on $\left(
\begin{array}{c}
 p   \\
1   \\
\end{array}
\right)$. The matrix $A$ is the most often used form for affine registration in medical imaging. 
The inverse of the affine transform is given by
$$p= R^{-1}q - R^{-1}c.$$
This can be written in a matrix form as
\bq
\left(
\begin{array}{c}
 p    \\
1   \\
\end{array}
\right)
=
\left(
\begin{array}{cc}
 R^{-1} & -R^{-1}c    \\
 0 \cdots 0&   1   \\
\end{array}
\right) \left(
\begin{array}{c}
 q   \\
1   \\
\end{array}
\right).
\eq
We denote the matrix form of the inverse as
$$A^{-} = \left(
\begin{array}{cc}
 R^{-1} & -R^{-1}c    \\
 0 \cdots 0&   1   \\
\end{array}
\right).$$

\index{estimation ! least squares}
\index{least squares estimation}

Let ${\bf p}_i$ be the $i$-th landmark  and its corresponding affine transformed points ${\bf q}_i$. Then we have
\bqn \left(\begin{array}{c}
 {\bf q}_i    \\
1   \\
\end{array}
\right)
= A 
\left(\begin{array}{c}
 {\bf p}_i    \\
1   \\
\end{array}
\right)
\label{eq:affine-LSE}.\eqn
Subsequently, we can estimate the affine transform matrix $A$ in the least squares fashion. The least squares estimation of $A$ is given by
$$\widehat{A} = \arg \min_{A \in \mathcal{G}}\sum_{i=1}^n \left\| \left(\begin{array}{c}
 {\bf q}_i    \\
1   \\
\end{array}
\right)
- A \left(\begin{array}{c}
 {\bf p}_i    \\
1   \\
\end{array}
\right)
 \right\|^2,$$
where $\mathcal{G}$ is an affine group. Since the direct optimization of this form is difficult, we alternately rewrite (\ref{affine-matrixform}) as 
\bq
\underbrace{\left(
\begin{array}{ccc}
{\bf q}_1 & \cdots &{\bf q}_n 
\end{array}
\right)}_{Q}
=
\left(
\begin{array}{cc}
 R &  c    \\
\end{array}
\right) 
\underbrace{
\left(
\begin{array}{ccc}
{\bf  p}_1& \cdots & {\bf p}_n\\
 1 &\cdots & 1
\end{array}
\right)}_{P}.
\eq
Then the least squares estimation is trivially given as
$$\left(
\begin{array}{cc}
 \widehat{R} & \widehat{ c}    \\
\end{array}
\right) 
= QP' (PP')^{-1}.
$$
This can be easily implemented in {\tt MATLAB}. Then the points ${\bf p}_i$ are mapped to $\widehat{R}{\bf p}_i + \widehat{c}$, which may not coincide with ${\bf q}_i$ in general.  

If we try to solve the least squares problem with an additional constraint, the problem can become complicated. For instance, if we restrict $R$ to be rotation only, i.e. $R'R = I$, iterative updates of least squares estimation are needed \citep{spath.2004}.

Affine transforms from the landmarks {\tt p} to {\tt q} is given computed as follows.

\begin{verbatim}
A =affine_transform(p',q')
F203_01_03a= affine_surface(A, F203_01_03);
figure; subplot(1,2,1);
figure_patch(F203_01_03a, [0.74 0.71 0.61], 0.5);
\end{verbatim}

The estimated affine transform is then applied to the landmark coordinates {\tt p}. The transformed landmarks are named as {\tt paffine} and displayed as follows.

\begin{verbatim}
paffine= p*A(1:3,1:3)' + repmat(A(1:3,4), 1, 24)';
hold on; 
scatter3(paffine(:,1),paffine(:,2),paffine(:,3),...
C, C, 'filled', 'b')
view([90 60]);  zoom(1.2); 
title('Affine transform of F203-01-03')
\end{verbatim}

Let us superimpose the affine transformed landmarks and the mandible on top of F155-12-08. As we can see, the affine transformed landmarks {\tt paffine} does not exactly match to {\tt q}. However, the distance between {\tt q} and {\tt paffine} is the smallest in the least squares fashion.

\begin{verbatim}
subplot(1,2,2);
figure_patch(F155_12_08, 'y', 0.5);
hold on; scatter3(q(:,1),q(:,2),q(:,3), C, C, 'filled', 'r')
hold on; figure_patch(F203_01_03a, [0.74 0.71 0.61], 0.5);
hold on; scatter3(paffine(:,1),paffine(:,2),paffine(:,3), ...
C, C, 'filled', 'b')
view([90 60]);  zoom(1.2); 
title('Landmark alignment')
\end{verbatim}

\section{Vector Data}
Vector image data are often obtained in connection with {\em deformation-based morphometry} (DBM) \citep{ashburner.2000,ashburner.2000b,chung.2001.ni} which does not require tissue segmentation in quantifying the shape of anatomy in MRI  \citep{ashburner.2000}. The 3D {\em displacement vector field} $u(x)$ is a vector map defined at each position $x$ that is required to move the anatomical structure at $x$ in an image $\mathcal{I}_1$  to the anatomically corresponding position in another image $\mathcal{I}_2$. The structure at $x$ in image $\mathcal{I}_1$ should match closely to the structure at $x+ u$ in image $\mathcal{I}_2$. The displacement vector field $u$ is usually estimated via volume- or surface-based nonlinear registration techniques. The $x$-, $y$- and $z$-components of displacement vector fields are usually stored in three separate Analyze files. We will study DBM in a later chapter. 

\begin{figure}[t!]
\centering
\includegraphics[width= 0.75 \linewidth]{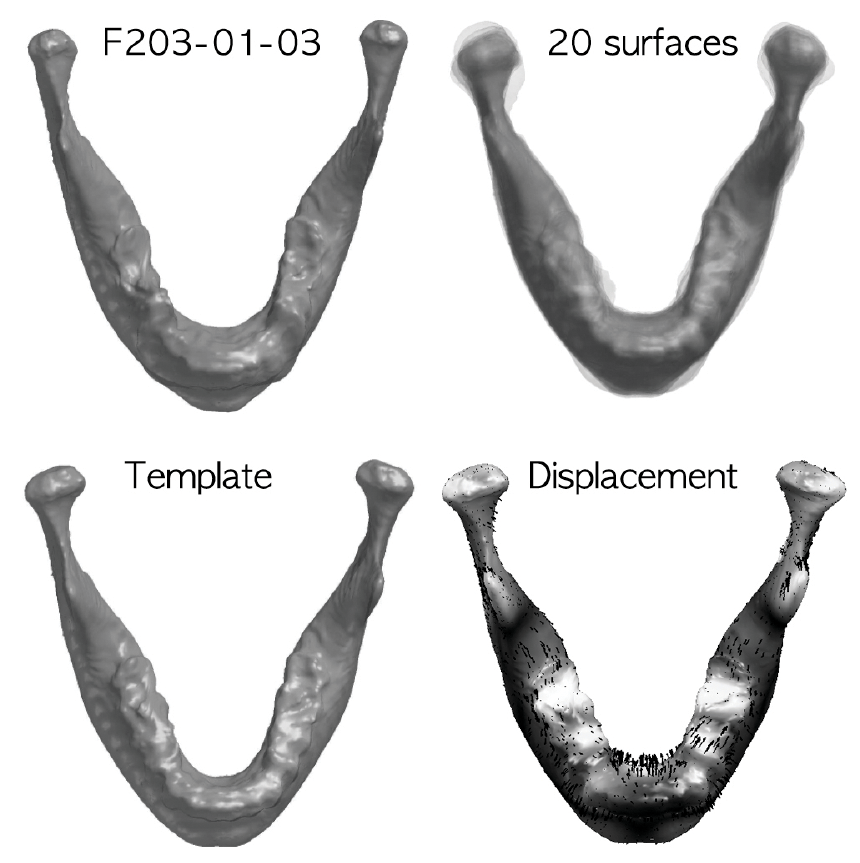} 
\caption{All 77 mandibles are affinely warped to F155-12-08 space. The affine transformed F203-01-03 surface is shown here. The superimposition of 20 affine registered mandibles shows local misalignment. Diffeomorphic registration is then performed to warp misaligned affine transformed mandibles. The surface meshes are then resampled in such a way that all the mandibles have identical mesh topology so that we can have anatomical correspondence across different mesh vertices. The average of deformation with respect to F155-12-08 provides the final population average template where statistical parametric maps will be constructed. The displacement vector from the template to F203-01-03 surface is shown as arrows.}
\label{fig:mandible-template}
\end{figure}

Here we briefly introduce the mandible surface deformation vector data set that will be used in the later chapter. {\tt mandible77subject.mat} contains surface deformation vector data for 77 subjects obtained from CT images. The data set contains the vertex coordinates {\tt vertices}, subject id {\tt id}, voxel size {\tt voxelsize},  mesh connectivity information {\tt faces} and affine transform matrices {\tt A}  for all 77 subject. Since the surface meshes are constructed from CT images with varying image resolution, it is necessary to keep track of the image resolution information. The surfaces are already affinely registered so there are no global size differences. In order to extract and display the mandible surface of the first subject as in Figure \ref{fig:mandible-template}, we run

\begin{verbatim}
load mandible77subjects.mat   

sub1.vertices=voxelsize(44)*vertices(:,:,1);
sub1.faces=faces;
figure;
subplot(2,2,1); figure_patch(sub1,[0.74 0.71 0.61],0.7);
view([90 60]); camlight
title(id(1))
\end{verbatim}

If we are interested in superimposing 20 surfaces (Figure \ref{fig:mandible-template}), we run

\begin{verbatim}
subplot(2,2,2);
for i=1:20
    i
    surf.faces=faces;
    surf.vertices=voxelsize(44)*vertices(:,:,i);
    hold on; figure_patch(surf,[0.74 0.71 0.61],0.1);
end;
view([90 60]); camlight;
title('20 surfaces')
\end{verbatim}

The template, which is the average of all 77 subjects, is given by averaging the surface coordinates in Figure \ref{fig:mandible-template}:

\begin{verbatim}
template.faces=faces;
template.vertices=voxelsize(44)*mean(vertices,3);
subplot(2,2,3); figure_patch(template,[0.74 0.71 0.61],0.7);
view([90 60]); camlight;
title('Template')
\end{verbatim}

\begin{verbatim}
disp = sub1.vertices - template.vertices;
displength=sqrt(sum(disp.^2,2));
mean(displength)

subplot(2,2,4); figure_surf(template,displength);
figure_quiver3surface(template, disp, 30);
view([90 60]); camlight
title('Displacement field')
colormap('hot')
\end{verbatim}

\section{Tensor and Curve Data}
Tensor and curve data are often obtained as the outcomes of diffusion tensor image (DTI) processing. DTI is a new imaging technique that has been used to characterize the macrostructure of biological tissues using magnitude, anisotropy and anisotropic orientation associated with water diffusion in the brain \citep{basser.1994}. DTI provides directional and connectivity information that MRI usually does not provide. The white matter fibers pose a physical constraint on the movement of water molecules along the direction of fibers. It is assumed that the direction of greatest diffusivity is most likely aligned to the local orientation of the white matter fibers \citep{mori.2002.nmr}. The directional information of water diffusion is usually represented
as a symmetric positive definite $3 \times 3$ matrix $D=(d_{ij})$ which is
usually termed as the {\em diffusion tensor} or {\em diffusion coefficients}.
The eigenvectors and eigenvalues of $D$ are obtained by solving $$D v = \lambda v,$$
 which results in 3 eigenvalues $\lambda_1 \geq \lambda_2 \geq \lambda_3$ and the corresponding eigenvectors $v_1, v_2, v_3$. The principal eigenvector $v_1$ determines the direction of the water diffusion, and is mainly used in streamline based tractography. Since there are exactly six unique elements in the diffusion tensor $D$, these six components of DTI are usually stored in six separate Analyze files. 
\index{diffusion tensor imaging (DTI)}
\index{imaging modalities ! diffusion tensor}

From DTI, we can obtain the fractional anisotropy (FA) map, which is defined as 
$$ {\tt FA} = \sqrt{ \frac{(\lambda_1 - \lambda_2)^2 + (\lambda_2 - \lambda_3)^2 + (\lambda_3 - \lambda_1)^2}{2(\lambda_1^2 + \lambda_2^2 + \lambda_3^2)}}.$$
FA value is a scalar measure between 0 and  1 and it measures the amount of anisotropicity of diffusion. 

\index{fractional anisotropy (FA)}
\index{file formats ! NIfTI-1}
\index{NIfTI-1 format}
\index{software package ! nii}

We have provided a sample FA-map, which is used as a study template in \citep{chung.2011.spie}. The FA-map is stored in NIfTI-1 file format and read using {\tt nii} package written by Jimmy Shen of Rotman Research Institute (\url{http://research.baycrest.org/~jimmy/NIfTI}). NIfTI-1 format is adapted from the widely used ANALYZE 7.5 file format (\url{http://nifti.nimh.nih.gov}).

\begin{figure}[t!]
\centering
\includegraphics[width=1\linewidth]{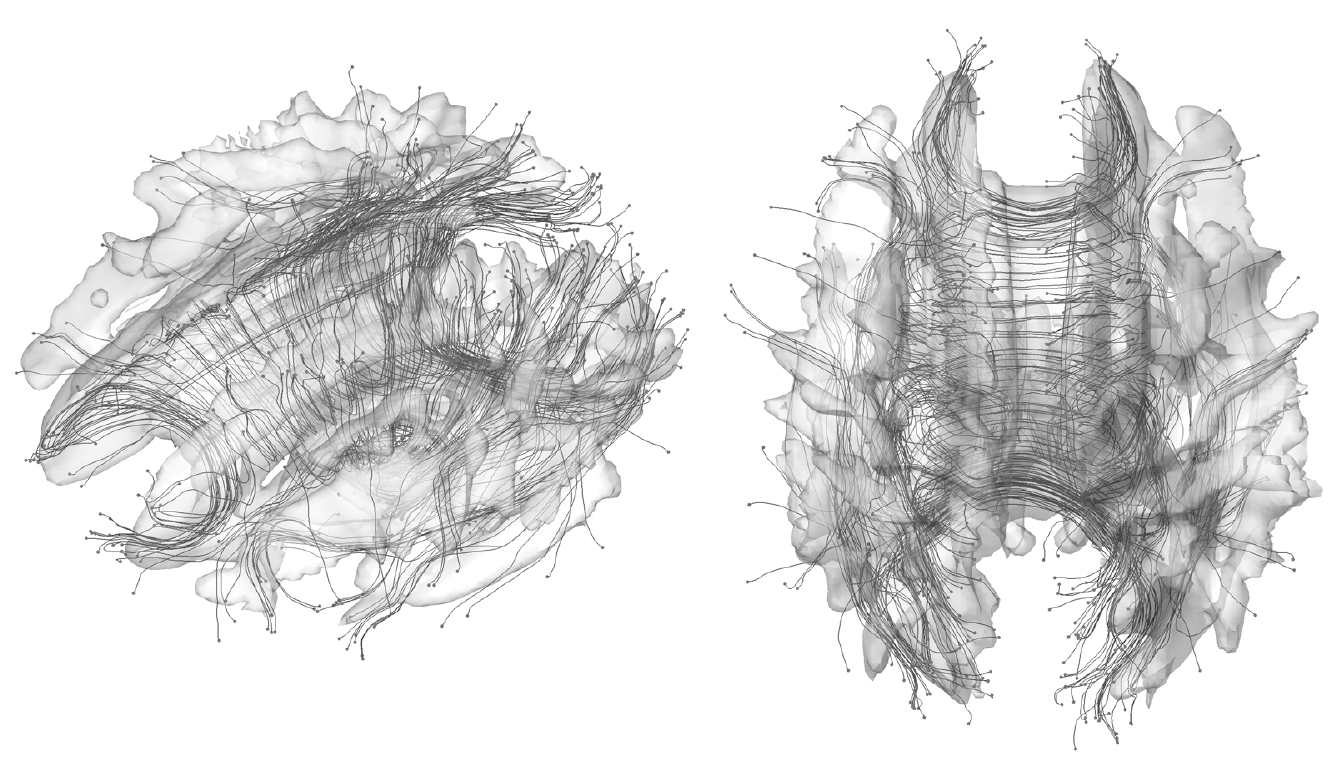}
 \caption{White matter fiber bundles obtained using TEND algorithm \citep{lazar.2003.hbm,cook.2006}.
The end points of tracts are identified and colored as red. The surface is the isosurface of the template FA map so most of tracts are expected to end  outside of the surface. The $\epsilon$-neighbor method uses the proximity of the end points in constructing the network graph.}\label{fig:endpoints}
\end{figure}
 
\begin{verbatim}
nii = load_nii('mean_diffeomorphic_initial6_fa.nii');  
d=nii.img;
surf = isosurface(d); 

coord=surf.vertices;
temp=coord;
temp(:,2)=coord(:,1);
temp(:,1)=coord(:,2);
surf.vertices=temp;

figure;
figure_patch(surf,[0.5 0.5 0.5],0.2);
axis off
set(gcf,'Color','w');
\end{verbatim}

We have to swap $x$- and $y$-coordinates of the image since the {\tt MATLAB} convention is slightly different from the usual brain imaging format.

DTI is usually processed again to obtain the streamline based tractography \citep{lazar.2003.hbm, cook.2006}. Figure \ref{fig:endpoints} shows the subsampled tractography result for a single subject. We will superimpose white matter fiber bundles stored in {\tt SL.mat} on top of the FA-map. {\tt SL.mat} contains 10000 white matter fiber tracts stored as the cell data structure. Since there are too many tracts, we subsample for every 30 tracts and plot them. 

\begin{verbatim}
load SL.mat

hold on;
for i=1:30:10000 
    tract=SL{i}';
    if size(tract,2)> 10
        hold on
        plot3(tract(1,:),tract(2,:),tract(3,:),'b')
    end;
end;
\end{verbatim}

\section{Brain Image Analysis Tools}
\index{software package ! SPM}
\index{software package ! AFNI}
\index{software package ! FSL}

Various neuroimage processing and analysis packages have been developed over the years. The main depository of neuroimaging tools is \url{www.nitrc.org}. The SPM (\url{www.fil.ion.ucl.ac.uk/spm}), AFNI  (\url{afni.nimh.nih.gov}) and FSL  (\url{www.fmrib.ox.ac.uk/fsl}) packages have been mainly designed for the whole brain volume based processing and univariate linear model type of analyses. The traditional statistical inference is then used to test hypotheses about the parameters of the model parameters. Although SPM and AFNI are probably the two most widely used analysis tools, their analysis pipelines are mainly based on a univariate general linear model and they do not have a routine for a multivariate or more complex statistical analysis. They also do not have the subsequent routine for correcting multiple comparison corrections for the multivariate linear models yet.

Unlike 3D whole brain volume based tools such as SPM, AFNI and FSL there are few cortical surface based tools such as the surface mapper (SUMA) \citep{saad.2004} and FreeSurfer (\url{surfer.nmr.mgh.harvard.edu}).
FreeSurfer is a widely used cortical surface segmentation and analysis tool. It consists of various image processing and segmentation tools. Also the spherical harmonic modeling tool SPHARM-PDM (\url{www.ia.unc.edu/dev/download/shapeAnalysis}) is available. However, these surface tools mainly do image processing and mesh representation and do not have the support for advanced statistical analyses. 

For advanced multivariate linear modeling, for instance, one has to actually use statistical packages such as Splus (\url{www.insightful.com}), R (\url{www.r-project.org}) and SAS (\url{www.sas.com}). These statistical packages do not interface with imaging data easily so the additional processing step is needed to read and write imaging data within the software. Further these tools do not have the random field based multiple comparison correction procedures so the users will likely export statistics maps to SPM or fMRISTAT (\url{www.math.mcgill.ca/keith/fmristat}) increasing the burden of additional processing steps. 
\index{software package ! fMRISTAT}

\subsection{\tt SurfStat}
\index{software package ! SurfStat}
SurfStat package (\url{www.stat.uchicago.edu/~worsley/surfstat}) was developed by the late Keith J. Worsley of McGill University utilizing a model formula and avoids the explicit use of design matrices and contrasts, which tend to be a hinderance to most end users not familiar with such concepts. Probably this is the most sophisticated statistical tool for analyzing brain images. SurtStat can import MNI \citep{macdonald.2000}, FreeSurfer (\url{surfer.nmr.mgh.harvard.edu})  based cortical mesh formats as well as other volumetric image data. The model formula approach is implemented in many statistics packages such as Splus (\url{www.insightful.com}), R (\url{www.r-project.org}) and SAS (\url{www.sas.com}). 
These statistics packages accept a linear model like
$$ {\tt P} =  {\tt Group} + {\tt Age} + {\tt Brain} $$
as the direct input for linear modeling avoiding the need to explicitly state the design matrix. {\tt P} is a $n \times 3$ matrix of coordinates, {\tt Age} is the age of subjects, {\tt Brain} is the total brain volume of subject and {\tt Group} is the categorical group variable (0=control, 1 = autism). This type of model formula has yet to be implemented in widely used SPM or AFNI  packages. 

The other major novelty of the SurfStat package is the inclusion of mixed-effects models that can explicitly model the within-subject correlation of image scans of the same subject. SurfStat package is therefore better suited for longitudinally collected study designs than other packages.

\subsection{Public Image Database}

Other than the data sets we provided in the book, a large longitudinal image dataset is also available to the public. Open Access Series of Imaging Studies (OASIS, \url{www.oasis-brains.org}) and Alzheimer's Disease Neuroimaging Initiative (ADNI, \url{www.loni.ucla.edu/ADNI}) are two widely distributed data sets. The ADNI was initiated in 2003 by the National Institute on Aging (NIA) and  the National Institute of Biomedical Imaging and Bioengineering (NIBIB) to determine if longitudinally collected MRI, PET and neuropsychological measurements  can predict the progression of mild cognitive impairment (MCI) and Alzheimer's disease (AD) (\url{www.adni-info.org}). There is also a Simulated Brain Database (SBD) (\url{brainweb.bic.mni.mcgill.ca/brainweb}). The SBD contains a set of realistic MRI data volumes produced by an MRI simulator. These data can be used by the neuroimaging community to evaluate the performance of various image analysis methods in a setting where the truth is known.

\section*{Acknowelgement}
This study was partially supported by NIH grants R01 EB022856 and R01 EB028753, and NSF grant MDS-2010778. 
\bibliographystyle{plainnat}
\bibliography{reference.2020.01.09}

\end{document}